\documentclass[epj]{svjour}
\def\bra#1{\langle{#1}|}
\def\ket#1{|{#1}\rangle}
\def\xiK{\xi_{\bf K}}
\def\HK{{\cal H}_{\bf K}}
\def\bracket#1#2{\langle{#1}|{#2}\rangle}

\usepackage{epsf}
\begin{document}
\title{A Non-Crossing Approximation for the Study of Intersite Correlations}
\author{Th.\ Maier\inst{1,2} \and M.\ Jarrell\inst{1} \and Th.\ Pruschke\inst{2} \and J.\ Keller\inst{2}
}                     
\offprints{Th. Maier}          
\institute{Department of Physics,
University of Cincinnati, Cincinnati, OH 45221 \and Institut f\"ur Theoretische Physik I,
Universit\"at Regensburg, 93040 Regensburg, Germany}
\date{Received: date / Revised version: date}
%
\abstract{
We develop a Non-Crossing Approximation (NCA) for the effective cluster 
problem of the recently developed Dynamical Cluster Approximation (DCA).
The DCA technique includes short-ranged correlations by mapping the lattice problem onto a self-consistently embedded 
periodic cluster of size $N_c$.  It is a fully causal and systematic 
approximation to the full lattice problem, with corrections ${\cal{O}}(1/N_c)$ 
in two dimensions.  The NCA we develop is a systematic approximation with 
corrections ${\cal{O}}(1/N_c^3)$. The method will be discussed in detail and 
results for the one-particle properties of the Hubbard model are shown. 
Near half filling, the spectra display pronounced features including a pseudogap and non-Fermi-liquid behavior due to 
short-ranged antiferromagnetic correlations.
} 
\maketitle
\section{Introduction}
\label{intro}
One of the most challenging tasks in theoretical condensed matter physics 
is the description of strongly correlated electron systems. The Coulomb 
interaction between electrons plays a dominant role in these systems and 
usually strongly influences the electronic properties and the physics of 
the ground-state phase. The discovery of heavy Fermion systems in the late 
seventies \cite{HFreviews} and high-$T_c$ super-conductors ten years 
later \cite{high_Tc} has stimulated strong experimental and theoretical 
interest in this field.  However, despite a multitude of attempts to 
describe such strongly correlated electron
systems theoretically, a complete understanding of the observed rich
physics has not yet been accomplished. Even the simplest model for
strongly correlated electron systems, the Hubbard model (HM), 
must be considered unsolved in more than one dimension \cite{lieb_wu}
after almost forty years of intensive study. Exact diagonalization or 
quantum Monte-Carlo studies for two or three
dimensions are restricted to small lattice sizes and predictions for
the thermodynamic limit are maybe problematic \cite{dagotto_review}.

However, in the  limit of infinite dimensions $D=\infty$, correlated 
lattice models undergo a significant simplification. Their dynamics 
become purely local and therefore the lattice problem can be mapped 
onto a generalized single impurity Anderson model coupled to a host, which 
has to be determined self-consistently \cite{Bra_Mie,Janis,Jarr92,Geo_Kot}. 
The Dynamical Mean Field Approximation used in the context of real
materials thus assumes that only local dynamics are present. Despite
the neglect of nonlocal correlations, this method has been shown to
capture several important features of e.g.\ the Hubbard model
\cite{Jarr92,jarrell92,TP95}.  Nevertheless it has some significant shortcomings due
to the mapping onto a purely local model. For instance, it does not
include the effect of nonlocal correlations like antiferromagnetic
spin fluctuations on the one-particle properties, and is not capable 
of describing nonlocal order parameters.
However, both effects are believed to be especially important for a 
description of the High-$T_c$ materials.  Here the one-particle 
spectra have shadow bands due to short-ranged antiferromagnetic fluctuations,
preformed pseudogaps due to superconducting or spin 
fluctuations\cite{shadow,preformed0,preformed1,preformed2,preformed3}, and the
superconducting order parameter is of nonlocal ($d$) character.

In order to include these types of nonlocal dynamics into the theory
there have been several efforts to add so-called $\frac{1}{D}$
corrections to the DMFA \cite{Dongen,Schi_Ing,obermeier}. However,
these methods either experience causality problems \cite{Dongen,Schi_Ing} 
(because of the necessary inclusion of nonlocal Green functions in self-energy
diagrams which do not have a negative semidefinite imaginary part), or
are restricted to the calculation of a few moments of the
spectral function \cite{moments}.

These shortcomings do not apply to the Dynamical Cluster Approximation
(DCA). This method systematically includes nonlocal short-ranged 
correlations while preserving causality \cite{hettler1,hettler2}. 
The DCA is a scheme which maps the lattice problem onto a self-consistently 
embedded effective finite-size cluster model. Due to the
finite size of the cluster, nonlocal corrections to the local dynamics
can be systematically included as the cluster size increases. 
The basic idea of the DCA is to take into account nonlocal physics by
calculating the self energy at selected points $\bf K$ in the
Brillouin zone and consider the self energy at these points to
represent the self-energy in the surrounding of these points ${\bf K}+
{\bf k}^\prime$: $\Sigma({\bf K})\approx\Sigma({\bf K}+{\bf k}^\prime)$.
The theory then maps the lattice problem onto an effective finite-size
system with periodic boundary conditions coupled to an external bath 
and the resulting system is solved self-consistently. The DCA has two 
well defined limits: It recovers the DMFA as the cluster size goes to 
1 and becomes the exact solution for the model under consideration as 
the cluster size goes to infinity.

The use of the DCA as an approximation can be justified as long as the
momentum dependence of the self-energy of the real system is
weak. This is obviously realized in high spatial dimensions where a
coarse grid of $\bf K$-points should capture all the basically
short-ranged dynamics. In two- or three-dimensional systems
the approximation is more crude, but can be motivated
by the observation that the dominant structures in the one-particle
dynamics are generated by local renormalizations, while nonlocal
effects only lead to minor renormalizations of these structures. Note
that this assumption automatically inhibits studies very close to
phase transitions since there strong, long-ranged fluctuations must be 
expected. However, sufficiently away from phase boundaries correlated
systems indeed show only mild momentum dependence of the one-particle
self energy as compared to its frequency dependence. 

So far only Quantum Monte Carlo simulations and exact enumeration have 
been used to solve this problem of a cluster in an external 
bath \cite{hettler2,hettler1}. In the DMFA the NCA has successfully been 
applied to the effective single impurity 
problem \cite{pruschke,pruschke2,obermeier1,schmalian2,Maier}. In this paper 
we introduce an extended version of the NCA to solve the effective periodic 
cluster model of the DCA. 

The paper is organized as follows. First a short review of the DMFA is
given, which is reproduced by the DCA for a single site cluster. 
Then we provide a microscopic definition of the DCA in terms of its
Laue function, and rederive the DCA algorithm using Baym's $\Phi$
functional formalism.  We then define the effective cluster model onto
which the lattice system is mapped by the DCA. An extended version of 
the NCA applicable to an impurity cluster of arbitrary size is discussed 
in detail and finally results for the one-particle properties of the 
Hubbard model are shown and compared to corresponding results of the 
DMFA.

\section{Dynamical Mean Field Approximation}
\label{sec:1}
In this paper we consider the single-band Hubbard model described by the 
Hamiltonian
\begin{eqnarray}
H=\sum_{ij\sigma}t_{ij}c^\dagger_{i\sigma}c^{}_{j\sigma}+
U\sum_{i}n_{i\uparrow}n_{i\downarrow}\quad\mbox{,}
\label{HM}
\end{eqnarray}
where $c_{i\sigma}^\dagger$ ($c_{i\sigma}^{}$) creates (destroys) an electron at site $i$ with spin sigma and $n_{i\sigma}$ are the corresponding number operators.
Lattice models of this kind simplify significantly in infinite
dimensions, while retaining their full local dynamics. Metzner and 
Vollhardt\cite{metzner} showed that the necessary rescaling of the 
kinetic energy as $1/\sqrt{D}$ leads to a collapse of all nonlocal 
diagrams in a skeleton expansion for the self-energy.  Consequently 
the corresponding Baym-Kadanoff $\Phi$ functional can be expressed 
in terms of local quantities only. 

\begin{center}
\begin{figure}[htb]
\epsfxsize=3.25in 
\epsffile{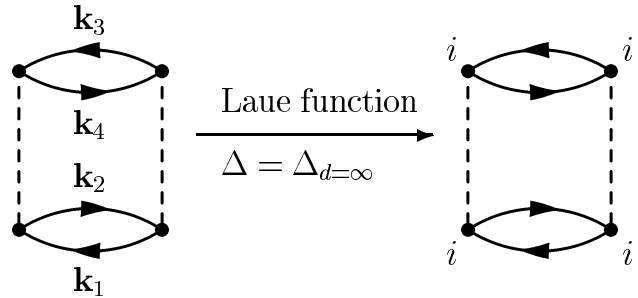}  

\vspace{0.5cm}
\caption{The Laue function (\ref{Laueinft}) leads to a collapse of the 
momentum dependence of the DMFA Baym-Kadanoff $\Phi_{DMFA}$ functional, 
illustrated for a second order contribution.}
\label{DMFA_Phi}
\end{figure}
\end{center}

M\"uller-Hart\-mann \cite{MueHa89} was able to deduce the same result 
by inspecting the momentum dependence of vertices in diagrammatic approaches as $D\rightarrow\infty$. 
For Hubbard-like models, the momentum dependence of each vertex in a
diagrammatic expansion of the functional $\Phi$ is completely characterized 
by the Laue function
\begin{eqnarray}
\Delta({\bf k}_1,{\bf k}_2,{\bf k}_3,{\bf k}_4)=\sum_{\bf r}e^{i({\bf k}_1-
{\bf k}_2+{\bf k}_3-{\bf k}_4)}\quad\mbox{,}
\end{eqnarray}  
where ${\bf k}_1$ and ${\bf k}_3$ (${\bf k}_2$ and ${\bf k}_4$) are the 
momenta entering (leaving) the vertex. In a conventional diagrammatic 
approach $\Delta({\bf k}_1,{\bf k}_2,{\bf k}_3,{\bf k}_4)=N\delta_{{\bf k}_1+
{\bf k}_3,{\bf k}_2+{\bf k}_4}$, which expresses momentum conservation on the 
vertex.  However as $D\to\infty$ M\"uller-Hartmann showed that the  Laue 
function reduces to
\begin{eqnarray}
\Delta_{D\rightarrow\infty}({\bf k}_1,{\bf k}_2,{\bf k}_3,{\bf k}_4)=
1+{\cal O}(1/D)\quad\mbox{.}
\label{Laueinft}
\end{eqnarray}
The DMFA assumes the same Laue function (\ref{Laueinft})  even in the 
context of finite dimensions. Therefore both the infinite-dimensional 
theory and the DMFA neglect momentum conservation at the internal vertices of 
irreducible diagrams and the momenta in the corresponding $\Phi_{DMFA}$ 
functional may be freely summed over the whole Brillouin zone. This leads to 
a collapse of the momentum dependent contributions to the functional 
$\Phi_{DMFA}$ and only local terms remain. This is illustrated in 
Fig.~\ref{DMFA_Phi} for a second order diagram.

The self-energy (a functional derivative of the functional $\Phi_{DMFA}$ 
with respect to a Green function leg) also becomes a functional of local 
propagators only and therefore becomes a constant in momentum space. 
Consequently the lattice problem can be mapped onto an effective 
impurity problem.

The DMFA has proven to capture the key features of strongly
correlated electron systems and to provide insight in the complicated
dynamics mediated by correlations. Despite its great success in the
description of correlated electron systems the DMFA has some
significant shortcomings due to the neglect of non-local dynamics. 

\section{Dynamical Cluster Approximation}
The DCA extends the DMFA through the inclusion of short-ranged dynamical 
correlations.  The DCA was introduced and discussed in detail in
\cite{hettler1,hettler2}. In this paper we will rederive the DCA
algorithm with an argument which is complimentary to that used by 
M\"uller-Hartmann \cite{MueHa89} to describe the DMFA.

\begin{center}
\begin{figure}[htb]
\epsfysize=2.25in 
\centering{\epsffile{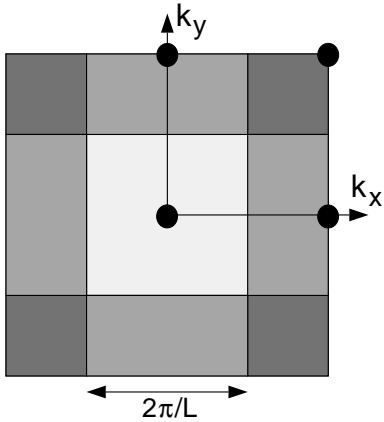}}  

\vspace{0.5cm}
\caption{The cluster momenta {\bf K} (filled circles) and coarse graining 
cells (different fill patterns) for a $N_c=4$ cluster in the Brillouin 
zone of a two dimensional lattice. The cells adjacent to the Brillouin zone 
boundary extend periodically to the opposite site.}
\label{BZ}
\end{figure}
\end{center}

\begin{center}
\begin{figure}[h]
\epsfxsize=3.25in 
\epsffile{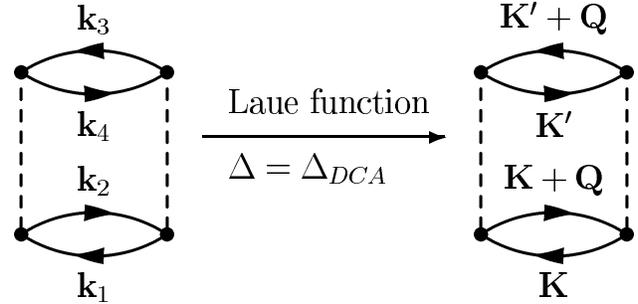}  

\vspace{0.5cm}
\caption{The DCA choice of the Laue function (\ref{LaueDCA}) leads to
  the replacement of the lattice propagators $G({\bf k}_1)$, $G({\bf
    k}_2)$, ... by coarse grained propagators $\bar{G}({\bf K})$,
  $\bar{G}({\bf K}^\prime)$, ... (Eq.~\ref{cgl1}) in the internal legs of $\Phi_{DCA}$, illustrated for a second order diagram.}
\label{DCA_Phi}
\end{figure}
\end{center}

The basic idea of the DCA is to partially restore the momentum
conservation relinquished by the DMFA. To this end the Brillouin-zone 
is divided into $N_c=L^D$ cells of size $(2\pi/L)^D$ (see Fig.~\ref{BZ}). 
Each cell is represented by a cluster momentum $\bf K$ in the center of 
the cell. We require that momentum conservation is (partially) observed 
for momentum transfers between cells, i.e. for momentum transfers larger 
than $\Delta k=2\pi/L$, but neglected for momentum transfers within a 
cell, i.e less than $\Delta k$. This requirement can be established by 
using the Laue function \cite{hettler2}
\begin{eqnarray}
\Delta_{DCA}=N_c\delta_{{\bf M}({\bf k}_1)+{\bf M}({\bf k}_3),{\bf M}({\bf k}_2)+{\bf M}({\bf k}_4)}\quad\mbox{,}
\label{LaueDCA}
\end{eqnarray}
where ${\bf M}({\bf k})$ is a function which maps ${\bf k}$ onto the cluster momentum ${\bf K}$ of the cell containing $\bf k$. With this choice of the Laue function the momenta of each internal leg in the corresponding functional $\Phi_{DCA}$ may be freely summed over the cell and each leg is replaced by the coarse grained average
\begin{eqnarray}
\bar{G}({\bf K})=\frac{N_c}{N}\sum_{\tilde{\bf k}}G({\bf K}+\tilde{\bf k})\quad\mbox{.}
\label{cgl1}
\end{eqnarray}
This is schematically illustrated in Fig.~\ref{DCA_Phi}.
The coarse grained Green function $\bar{G}({\bf K})$ and corresponding self-energy $\bar{\Sigma}({\bf K})$ 
\begin{eqnarray}
\bar{\Sigma}({\bf K})=\frac{N_c}{N}\sum_{\tilde{\bf k}}\Sigma({\bf K}+\tilde{\bf k})
\label{cgl2}
\end{eqnarray}  
 are functions of the cluster momenta ${\bf K}$ only. The $\tilde{\bf k}$ summation in (\ref{cgl1}) and (\ref{cgl2}) runs over the $N/N_c$ momenta of the cell about the cluster momentum $\bf K$ and $G$ ($\Sigma$) is the full lattice propagator (self-energy). 

With this definition of $\Delta_{DCA}$, the DCA estimate of the Baym-Kadanoff
functional $\Phi_{DCA}$ becomes \cite{hettler2}
\begin{equation}\label{phi}
\Phi_{DCA}(\bar{G})=\sum_l\frac{1}{2l}{\rm tr}\left[\bar{\Sigma}^{(l)}\bar{G}\right]
\end{equation}
where $\bar{\Sigma}^{(l)}$ is the set of irreducible, coarse grained
self-energy diagrams of $l^{\rm th}$ order in the interaction $U$ and
the trace indicates summation over frequency, cluster momenta and
spin.
The DCA result for the free energy of the lattice is
\begin{equation}
\Omega_{DCA}=-k_BT(\Phi_{DCA}-{\rm tr}\Sigma G-{\rm tr}\, 
{\rm ln}[-G])\quad\mbox{.}
\end{equation}
$\Omega_{DCA}$ is stationary with respect to the lattice Green function $G$
\begin{equation}
\frac{\delta {\Omega}_{DCA}}{\delta G}=k_BT[-\bar{\Sigma}+\Sigma]=0\quad\mbox{,}
\label{stationary}\,
\end{equation}
if $\bar{\Sigma}$ is taken as an approximation for the self-energy $\Sigma$ 
of the full lattice Green function $G$ (the left hand side of (\ref{stationary}) 
follows from $\delta\bar{G}_{\bf K}/\delta G_{\bf k}=\delta_{{\bf K},
{\bf M}({\bf k})}$ and $\delta\Phi_{DCA}/\delta\bar{G}=\bar{\Sigma}$). We 
have shown previously that $\Sigma=\bar{\Sigma}+{\cal{O}}(1/N_c)$ in two 
dimensions and includes dynamical intersite correlations of range 
$\pi/\Delta k=L/2$ \cite{hettler2}.

The coarse grained Green function (\ref{cgl1}) then takes the form
\begin{eqnarray}
\bar{G}({\bf K},z)=\frac{N_c}{N}\sum_{\tilde{\bf k}}
\frac{\displaystyle 1}{\displaystyle z-\epsilon_{{\bf K}+\tilde{\bf k}}+
\mu-\bar{\Sigma}({\bf K},z)}\quad\mbox{,}
\label{G_bar_sum}
\end{eqnarray}
where the self-energy at momentum ${\bf k}={\bf K}+\tilde{\bf k}$, 
$\Sigma({\bf k})$ is replaced by its coarse grained average 
$\bar{\Sigma}({\bf K})$ and $z=\omega+i\delta$.
Note that the choice of the coarse grained Green function
(\ref{G_bar_sum}) has two well defined limits with respect to the
cluster size $N_c$. For $N_c=1$ the $\tilde{\bf k}$ summation runs
over the entire first Brillouin-zone, $\bar{G}$ is the local Green
function, thus the DMFA algorithm is recovered. When $N_c=\infty$ the
$\tilde{\bf k}$ summation vanishes and the DCA becomes equivalent to
the exact solution of the Hubbard model. 

In order to apply the NCA to solve the effective cluster problem it 
is convenient to write the coarse grained Green function (\ref{G_bar_sum}) 
in a more suitable form.  We use the independence of the self-energy 
$\bar{\Sigma}({\bf K})$ on the integration variable $\tilde{\bf k}$ 
to write $\bar{G}$ in the form
\begin{eqnarray}
\bar{G}({\bf K},z)=\frac{\displaystyle 1}{\displaystyle z-\bar{\epsilon}_{\bf K}+\mu-\bar{\Sigma}({\bf K},z)-\Gamma({\bf K},z)}\quad\mbox{,}
\label{G_bar_loc}
\end{eqnarray}
where ${\rm Im}\Gamma({\bf K})<0$ (see Appendix \ref{app:a}).
This is just the Green function of an effective cluster model with periodic 
boundary conditions coupled to a Fermionic bath described by the host 
function $\Gamma({\bf K})$. Hence we can obtain the coarse grained self 
energies $\bar{\Sigma}({\bf K})$ by solving a generalized cluster model.

The DCA cluster problem may then be solved by iteration. The iteration loop 
starts with a guess for the initial cluster self-energy. By computing the 
coarse grained lattice Green function (\ref{G_bar_sum}) we get the input 
quantities for the effective cluster model. The effective dispersion 
$\bar{\epsilon}_{\bf K}$ is given by the average 
\begin{eqnarray}
\bar{\epsilon}_{\bf K}=\frac{N_c}{N}\sum_{\tilde{\bf k}}\epsilon_{{\bf K}+
\tilde{\bf k}}
\label{eK_bar}
\end{eqnarray}
and the cluster electrons are coupled to the host function $\Gamma({\bf K})$. 
This will be described in more detail in section \ref{ICL} and \ref{sec:NCA}.
Given the effective dispersion $\bar{\epsilon}_{\bf K}$ and the host
$\Gamma({\bf K})$ the interacting Green function $G_{c}({\bf K})$ of
the effective cluster model can be calculated by some suitable method. 
The cluster self energy is then obtained via 
$\Sigma_c({\bf K},z)=
z-\bar{\epsilon}_{\bf K}+\mu-\Gamma({\bf K},z)-G_{c}^{-1}({\bf K},z)$ 
and the iteration closes by calculating a new $\bar{G}({\bf K})$ with 
Eq.~\ref{G_bar_sum}. This procedure is repeated until 
$G_{c}({\bf K})=\bar{G}({\bf K})$ within the desired accuracy.

\section{Effective Cluster Model \label{ICL}}
To solve the cluster problem with the NCA, we must first define a 
Hamiltonian for the cluster. The parameters of the Hamiltonian are given
by the Green function (\ref{G_bar_loc}). The corresponding cluster 
Hamiltonian 
\begin{eqnarray}
H_{cluster}=H_{loc}+H_{med}
\label{H_cluster}
\end{eqnarray}
is most conveniently constructed in momentum space. Its local part is given by 
\begin{eqnarray}
H_{loc}&=&\sum_{{\bf K}\sigma}\bar{\epsilon}_{\bf K}f^\dagger_{{\bf K}\sigma}f^{}_{{\bf K}\sigma}\nonumber\\
&+&\frac{U}{N_c}\sum_{\stackrel{{\bf K},{\bf K}^\prime}{\bf Q}}f^\dagger_{{\bf K}+{\bf Q}\uparrow}f^{}_{{\bf K}\uparrow}f^\dagger_{{\bf K}^\prime -{\bf Q}\downarrow}f^{}_{{\bf K}^\prime \downarrow}\quad\mbox{,}
\label{H_loc}
\end{eqnarray}
where $f^\dagger_{{\bf K}\sigma}$ ($f^{}_{{\bf K}\sigma}$) creates
(destroys) an electron with momentum $\bf K$ and spin $\sigma$. $U$
is the local Coulomb repulsion for two electrons residing on the same
cluster site. Since this interaction is local, it is unchanged by the
coarse-graining procedure\cite{hettler2}.  Note that the effective 
dispersion $\bar{\epsilon}_{\bf K}$ 
of the cluster states is given by the average bare dispersion in
the cell (\ref{eK_bar}). The coupling of the local cluster states with the 
host has the form
\begin{eqnarray}
H_{med}&=&\frac{1}{\sqrt{N}}\sum_{{\bf K},{\bf k}^\prime \sigma} V_{{\bf K},{\bf k}^\prime}
(f^\dagger_{{\bf K}\sigma}c^{}_{{\bf K}+{\bf k}^\prime \sigma}+h.c.)\nonumber\\&+&\sum_{{\bf k}\sigma}\varepsilon_{\bf k}c^\dagger_{{\bf k}\sigma}c^{}_{{\bf k}\sigma}\quad\mbox{,}
\label{H_med}
\end{eqnarray}
where $c^\dagger_{{\bf K}+{\bf k}^\prime}$ ($c^{}_{{\bf K}+{\bf k}^\prime}$) describe the effective medium in terms of free fermions with a dispersion $\varepsilon_{\bf k}$. Note that in contrast to the single impurity model, the local states given by $f^\dagger_{{\bf K}\sigma}$ couple only to fermions with momenta ${\bf k}={\bf K}+{\bf k}^\prime$ within the cell about the cluster momentum ${\bf K}$. Therefore the corresponding hybridization function
\begin{eqnarray}
\Gamma^\prime({\bf K},z)=\frac{1}{N}\sum_{{\bf k}^\prime}\frac{\displaystyle |V_{{\bf K},{\bf k}^\prime}|^2}{\displaystyle z-\varepsilon_{{\bf K}+{\bf k}^\prime}}
\label{sigma_V}
\end{eqnarray}
becomes $\bf K$-dependent and the interacting cluster Green function
finally reads 
\begin{eqnarray}
G_{c}({\bf K},z)=\frac{\displaystyle 1}{\displaystyle z-\bar{\epsilon}_{\bf K}+\mu-\Sigma_c({\bf K},z)-\Gamma^\prime({\bf K},z)}\quad\mbox{.}
\label{G_f}
\end{eqnarray}
$\Sigma_c({\bf K},z)$ denotes the proper one-particle self energy effects due to the local Coulomb repulsion between the $f$-electrons.

Comparing this result with the coarse-grained Green function of the lattice
(\ref{G_bar_loc}) one finds identical structures provided that the cluster 
self-energy $\Sigma_c({\bf K})$ equals that of the lattice 
$\bar{\Sigma}({\bf K})$, and that
\begin{eqnarray}
\Gamma^\prime({\bf K})=\Gamma({\bf K})\,.
\label{S_V_eq_D}
\end{eqnarray}
The latter substitution ensures that the solution of the effective 
cluster model is also the solution of the coarse-grained lattice
problem discussed in the last section.  

\section{Extended Version of the Non Crossing Approximation\label{sec:NCA}}
In the following we show how to solve the effective cluster model with an 
extended version of the NCA.  The effective cluster model is defined by 
the Hamiltonian (\ref{H_cluster})  with the effective medium fixed by 
Eq.~\ref{S_V_eq_D}.  The NCA is a perturbational expansion around the
molecular limit, i.e. it starts with the eigenstates of the local part
$H_{loc}$ of the cluster Hamiltonian (\ref{H_cluster}). The expansion
is performed with respect to the coupling to the effective medium
$H_{med}$, where the quasi-free fermions are described by the host
function $\Gamma({\bf K})$, see Eq.~\ref{S_V_eq_D}. The Fermionic
operators defined on the cluster are expanded in terms of the Hubbard
operators $X_{mn}=\left|m\right>\left<n\right|$, e.g.
\begin{eqnarray}
f^{}_{{\bf K}\sigma}=\sum_{m,n}F^{{\bf K}\sigma}_{mn}X_{mn}\quad\mbox{,}
\label{expansion}
\end{eqnarray}
where $\{\left|m\right>\}$ are the eigenstates of the local Hamiltonian 
\begin{eqnarray}
H_{loc}=\sum_m E_m X_{mm}
\label{H_loc_m}
\end{eqnarray}
with eigenenergies $E_m$ and $F^{{\bf K}\sigma}_{mn}=
\langle m|f^{}_{{\bf K}\sigma}|n\rangle$. The hybridization 
term (\ref{H_med}) becomes 
\begin{eqnarray}
H_{med}&=&\frac{1}{\sqrt N}\sum_{{\bf K},{\bf k}^\prime \sigma}
\sum_{m,n} V_{{\bf K},{\bf k}^\prime}(c^\dagger_{{\bf K}+{\bf k}^\prime}F^{{\bf K}\sigma}_{mn}X_{mn}+h.c.)\nonumber\\&+&\sum_{{\bf k}\sigma}\varepsilon_{\bf k}c^\dagger_{{\bf k}\sigma}c^{}_{{\bf k}\sigma}\quad\mbox{.}
\label{H_med_m}
\end{eqnarray}
Since the Hubbard-operators do not obey stan\-dard Ferm\-ion\-ic or Bo\-sonic
commutation relations, the conventional Feynman diagram technique
cannot be used for a perturbation expansion and the concept of
resolvents must be introduced instead \cite{keiter}. Their matrix-elements in 
the space of the local eigenstates
have the form
\begin{eqnarray}
[\hat{R}^{-1}]_{mn}(z)=(z-E_m)\delta_{mn}-\hat{\Sigma}_{mn}(z)\quad\mbox{.}
\label{P_m}
\end{eqnarray}
In general non-diagonal elements $\hat{R}_{mn}$ of the resolvent exist, 
but if the hybridization term $H_{med}$ does not break the symmetry of 
the cluster local Hamiltonian $H_{loc}$ they are zero. $\hat{\Sigma}$
describes self-energy effects due to the hybridization with the
effective medium. Note that $\hat{\Sigma}$ collects the
renormalizations of the individual local states $\{|m\rangle\}$ and must
not be confused with the proper one-particle self energy of the
cluster, $\bar{\Sigma}({\bf K},z)$.

\begin{center}
\begin{figure}[h]
\epsfxsize=3.25in 
\epsffile{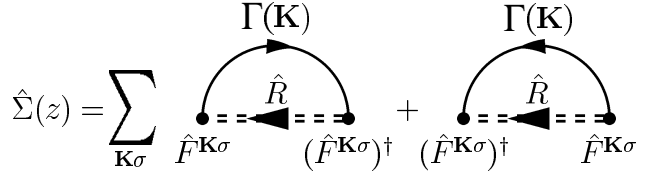}  
\caption{NCA self-energy of the resolvent $\hat{R}$ due to the coupling of 
the local cluster states to the host $\Gamma({\bf K})$.}
\label{NCA_self}
\end{figure}
\end{center}

In the NCA the self-energy matrix $\hat{\Sigma}$ is obtained by calculating 
the two diagrams illustrated in Fig.~\ref{NCA_self}, which correspond to
\begin{eqnarray}
\hat{\Sigma}(z)&=&-\frac{1}{\pi}\sum\limits_{{\bf K}\sigma}\left[\int\limits_{-\infty}^{+\infty}\!d\varepsilon\,f(\varepsilon){\rm Im}\Gamma({\bf K},\varepsilon)\hat{F}^{{\bf K}\sigma}\hat{R}(z+\varepsilon)(\hat{F}^{{\bf K}\sigma})^\dagger\right.\nonumber\\
&+&\left.\int\limits_{-\infty}^{+\infty}\!d\varepsilon\,f(-\varepsilon){\rm Im}\Gamma({\bf K},\varepsilon)(\hat{F}^{{\bf K}\sigma})^\dagger\hat{R}(z-\varepsilon)\hat{F}^{{\bf K}\sigma}\right]\,\mbox{.}
\label{self_m}
\end{eqnarray}
The coupled singular integral equations (\ref{P_m}) and (\ref{self_m}) have 
to be solved self-consistently. 

        Higher order corrections to these equations come in the form of vertex 
corrections or crossing diagrams. For the orbitally non-degenerate single 
impurity Anderson model it is well known that to obtain the correct value 
for the low-energy scale one has to sum all diagrams up to fourth order in 
the coupling $V$\cite{pru89}. For a cluster of size $N_c\ge1$ and finite
value for $U$, this requirement would mean that one must include vertex 
corrections.  From our former experience with vertex corrections for the 
single impurity case ($N_c=1$) we expect a strong renormalization of
low-energy scales. However, for high energy features in the spectra or high 
temperature properties like magnetism the vertex corrections yield only 
negligible effects. They also do not affect general local properties like 
universality or scaling\cite{fischer}.    

In the present context, the hybridization strength is not an adjustable 
parameter, so it does not make sense to use it to classify the higher-order
corrections. In fact, both the effective hybridization strength between the 
cluster and its host, and the degeneracy and magnitude of the cluster states 
depend upon $N_c$.  Therefore, a far more important expansion parameter is the 
inverse cluster size $1/N_c$.  Since the eigenenergies $E_m$ of the cluster 
scale as $N_c$, the resolvent behaves like $\hat{R}\sim{\cal O}(1/N_c)$. 
Taken together the sum over the cluster momenta $\bf K$ and the resolvent in 
(\ref{self_m}) are of order one. Thus the 
$N_c$-dependence of the NCA-self energy matrix $\hat{\Sigma}$ is determined 
by the $N_c$-dependence of $\Gamma$.  We show in Appendix \ref{app:b} that 
the host function $\Gamma$ is of order ${\cal O}(1/N_c)$. Therefore for 
$N_c\rightarrow\infty$ the NCA-self energies vanish - as expected since 
the coupling to the host vanishes - and the cluster problem is solved 
exactly by the eigenstates $\{|m\rangle\}$.  In the form (\ref{self_m}) the 
NCA equations are exact up to the second order in the coupling (\ref{H_med}), 
i.e. first order in $\Gamma \sim {\cal O}(1/N_c)$. 

\begin{center}
\begin{figure}[h,t]
\epsfxsize=3.25in 
\epsffile{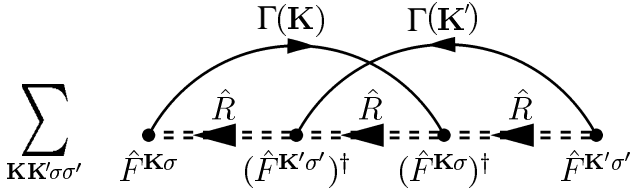}  
\caption{Leading order crossing diagram is of order ${\cal O}(1/N_c^3)$, since 
it involves five quantities of order ${\cal O}(1/N_c)$ but only two sums over 
the cluster momenta $\bf K$.} 
\label{cross}
\end{figure}
\end{center}

To estimate the role of vertex corrections we show
in Fig.~\ref{cross} one of the leading order 
corrections to the NCA-self energy (\ref{self_m}).  This crossing diagram 
involves two $\Gamma$-lines and three resolvents $\hat{R}$, 
but only two sums over the cluster momenta $\bf K$, therefore this 
diagram is of order ${\cal O}(1/N_c^3)$.  In fact all crossing 
diagrams are of this order or higher.  Hence for $N_c\rightarrow\infty$ 
the NCA algorithm becomes exact with corrections ${\cal O}(1/N_c^3)$. We 
are thus confident that at least the qualitative aspects of
our results will be unaffected by higher order diagrams. Since on the other 
hand an inclusion of vertex corrections is associated with a tremendous
numerical effort, we refrain from taking them into account for the
time being.
With the expansion (\ref{expansion}) the cluster Green function
$G_{c}=\langle\langle f^{}_{{\bf K}\sigma};f^\dagger_{{\bf
    K}\sigma}\rangle\rangle$ can be written in terms of the Hubbard
operators as
\begin{eqnarray}
G_{c}({\bf K},z)=\sum\limits_{\stackrel{mn}{m^\prime n^\prime}} 
F^{{\bf K}\sigma}_{mn}F^{{\bf K}\sigma}_{n^\prime m^\prime}\langle\langle X_{mn};X_{m^\prime n^\prime}\rangle\rangle_z\;\;.
\label{G_imp}
\end{eqnarray}
Within the NCA, the correlation function on the right-hand side of (\ref{G_imp})
can be written as
\begin{eqnarray}
\langle\langle X_{mn};X_{m^\prime n^\prime}\rangle\rangle_\omega=
\frac{\displaystyle 1}{\displaystyle \bar{Z}}\int\limits_{-\infty}^{+\infty}\!&d\varepsilon&\;e^{-\beta\varepsilon}(\rho_{n^\prime m}(\varepsilon)\hat{R}_{nm^\prime}(\varepsilon+\omega)\nonumber\\
&-&\rho_{nm^\prime}(\varepsilon)\hat{R}_{n^\prime m}(\varepsilon-\omega))\;\mbox{.}
\label{Xmm}
\end{eqnarray}
Here $\rho_{nm}$ is the spectral density of the resolvents,
$\rho_{nm}=-\frac{1}{\pi}{\rm Im}\hat{R}_{nm}(\omega)$ and 
$\bar{Z}=Tr\int\limits_{-\infty}^{\infty}\!d\varepsilon\,
e^{-\beta\varepsilon}\hat{\rho}(\varepsilon)$ 
is the cluster partition function.

\section{Results}

The DCA enables us to include
short-ranged nonlocal dynamical correlations neglected in the DMFA. The 
main goal of this section will be to show that this is indeed the case 
and to present some systematics on how these nonlocal correlations evolve 
and in what way their influence depends on system parameters like filling 
and band structure. To this end we present results for the Hubbard 
model (\ref{HM}) on a square lattice in two dimensions. For nearest-neighbor 
hopping the dispersion is $\epsilon_{\bf k}=-2t(\cos k_x+\cos k_y)$, 
i.e.\ the bandwidth of the noninteracting system $W=8t$.  Calculations were 
performed for a 1x1 cluster ($N_c=1$), which is equivalent to a DMFA 
calculation, and for a 2x2 cluster ($N_c=4$). A comparison of the results 
for both cluster sizes is used to study the effect of nonlocal correlations 
present in the $N_c=4$, but neglected in the $N_c=1$ calculation.  

The total number of cluster eigenstates scales with the cluster size
$N_c$ like $4^{N_c}$. The large number of eigenstates (256) for the 2x2
cluster results in an expensive numerical calculation.  The complexity 
of the problem can be reduced by taking into account 
the symmetries of the cluster Hamiltonian (\ref{H_cluster}).  Since our 
studies are restricted to the paramagnetic phase we can drop the spin index 
due to the SU(2) symmetry of the cluster Hamiltonian (\ref{H_cluster}). 
A further reduction in complexity can be achieved by using the point-group
symmetry of the cluster.  However, this depends strongly on the choice
of the cluster momenta $\bf K$ within the first Brillouin zone. A
priori, there is no restriction in the choice of the cluster momenta
$\bf K$ within the first Brillouin-zone, since in the derivation of
the DCA algorithm no special assertion about the cluster $\bf K$
points was made. One  e.g.\ could choose all $\bf K$ momenta to lie on
the Fermi surface. However, to identify eigenstates which are
degenerate due to the geometric symmetry, one has to classify the
eigenstates according to the cluster momenta $\bf K$. Since the
cluster Hamiltonian (\ref{H_cluster}) conserves the cluster momentum, its
many-particle eigenstates can be classified according to their total
momentum, which is just the sum of the momenta of the participating one 
particle states. This approach restricts the freedom in choosing the cluster
momenta ${\bf K}$ to exactly one possibility. The only set of cluster
momenta $\bf K$ which form a group under addition is $K_{\alpha l}=l\pi$, 
where $l=0,1$ and $\alpha=x$ or $y$. This set corresponds
to periodic boundary conditions for the cluster in real space. With
this choice of the cluster momenta we are able to classify the
eigenstates according to their total particle number, total momentum,
total spin and their z-component of the spin.  The degeneracy in the 
cluster momentum points $(0,\pi)$ and $(\pi,0)$ and the spin symmetry 
finally lead to an effective number of 123 non-degenerate eigenstates 
which have to be considered.  Then effectively only resolvents with 
different energies occur in our calculations. 

The remaining  numerical task of calculating the coupled equations 
(\ref{P_m}) and (\ref{self_m}) self-consistently becomes for\-mi\-dable as 
the cluster size increases.  Although the study of larger cluster sizes 
is in-principle possible\cite{ED}, presently this restricts our calculations 
to a cluster size of $N_c=4$.   Also the evaluation of two-particle
correlation functions is formally possible, but the associated numerical 
effort scales much worse with the cluster size $N_c$ than calculations on 
the one particle level. Hence our calculations are currently limited to 
one-particle Green functions.  

We will present results for local single-particle spectra as well as for 
the bandstructure. Since within the DCA we calculate the self-energy at 
the selected momenta $\bf K$ only, we need to perform a bilinear 
interpolation of the self energy between the cluster momenta $\bf K$ to 
calculate nonlocal spectra. We also show results for the self-energy at 
the Fermi-surface. The shape of the Fermi-surface 
is not a priori clear. In order to evaluate the Fermi surface we take 
the bilinear interpolated self-energy and calculate the occupation 
$n(\bf k)$ in momentum space along various directions in the Brillouin 
zone. The maximum value of $|{\rm d}n({\bf k})/{\rm d}{\bf k}|$ along 
these directions then marks the Fermi-surface and we get the self-energy 
at the Fermi surface from the interpolated form.

In the following we will concentrate at first on a generic set of values 
for the temperature $T$ and Coulomb parameter $U$, namely $T=W/15$ and
$U=W/2$. These values for $U$ and $W$ assure that for the half filled
case the system is metallic and far from the Mott-Hubbard transition, which 
is expected to occur at values $U\approx W$.
This choice allows us to directly compare the 
properties of the expected metallic phases at and off half filling.   The 
equally interesting question, of how the Mott-Hubbard transition at half 
filling will be affected by nonlocal correlations is left out for the time
being and will be the subject of a forthcoming publication.
\begin{center}
\begin{figure}[h]
\epsfxsize=3.25in 
\epsffile{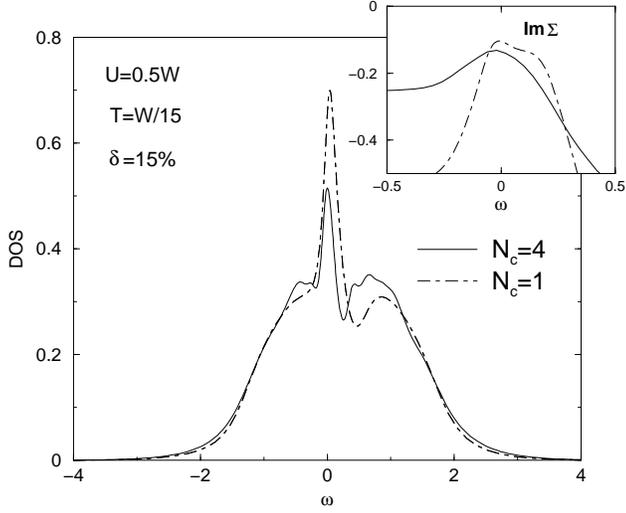}  
\caption{Local density of states for the 15\% doped 1x1 ($N_c=1$) and 2x2 
($N_c=4$) cluster at fixed temperature $T=W/15$ and interaction $U=0.5W$. 
Inset: Imaginary part of the corresponding self energies in a narrow region 
around $\omega=0$.}
\label{res1}
\end{figure}
\end{center}
Fig.~\ref{res1} shows the density of states for both the 1x1 and 2x2
clusters for a doping of $\delta=15\%$. Both spectra display qualitatively 
similar features, namely the typical Hubbard bands and an enhanced density 
of states at the Fermi level $\omega=0$. For both cluster sizes the 
imaginary part of the corresponding self energies shows  a parabolic 
minimum at the Fermi level as expected for a Fermi liquid, where 
$\left.-{\rm Im}\Sigma(\omega)\right|_{\omega\rightarrow 0}
\propto\omega^2+\pi T^2$. 
In the case of the 2x2 cluster the self energy at the Fermi wave vector 
$\bar{\Sigma}(k_F,\omega)$ is obtained from the above mentioned bilinear  
interpolation. Here $k_F$ lies along the diagonal from $(0,0)$ to $(\pi,\pi)$;
however, for this set of parameters the self energy on the Fermi surface 
is almost isotropic and does not change its qualitative behavior as a 
function of the wave vector on the Fermi surface. The influence of nonlocal 
short-ranged dynamical correlations is visible in the 2x2 cluster 
calculation only in a slightly enhanced scattering rate at the Fermi 
level and therefore in a slightly reduced density of states as 
compared to the 1x1 result. The additional
structures on top of the Hubbard bands can be traced back to the
complex multiplet structure of the cluster. However, the qualitative 
physics remains unchanged by increasing the cluster size. This observation is
consistent with the fact that at such strong doping antiferromagnetic
fluctuations have practically died out and should thus show no
significant influence on the physics of the system. The appearance of
the quasi-particle resonance at low enough temperatures is well known
in the case of the 1x1 cluster (DMFA): There it was shown that the
evolution of this quasi-particle resonance with decreasing temperature
is accompanied by a reduction of the effective local magnetic moment
\cite{jarrell92,AdvPhy}. This interplay of both effects is a
fingerprint of the Kondo effect occurring in the single impurity
Anderson model, which underlies the DMFA. Our results thus suggest
that for the lattice system the physics is quite similar and the
quasi-particle resonance at the Fermi level reflects Kondo like
physics. It is important to note that this means that the Kondo like
behavior in the Hubbard model is, at least for moderately to strongly
doped systems, a real feature of the model and not an artifact of the
limit of large dimensions.

For weakly doped or half filled systems, short-ranged antiferromagnetic spin 
fluctuations will be present and strong even at 
temperatures well above a magnetic transition. One thus expects
that physics of the system will be strongly
influenced and may even develop non-Fermi-liquid-like behavior.
\begin{center}
\begin{figure}[h]
\epsfxsize=3.25in 
\epsffile{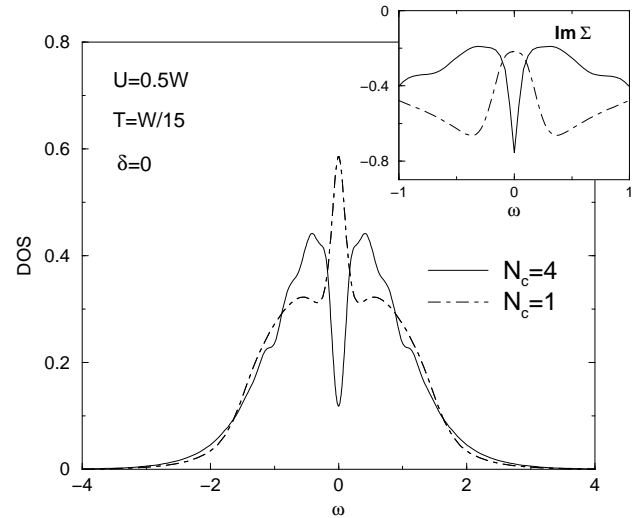}  
\caption{Local density of states for the  1x1 ($N_c=1$) and 2x2 ($N_c=4$) 
cluster at fixed temperature $T=W/15$, interaction $U=0.5W$ and half 
filling. Inset: Imaginary part of the corresponding self energies in a 
narrow region around $\omega=0$.}
\label{res2}
\end{figure}
\end{center}
Since these fluctuations will be strongest in the extreme case of half 
filling, we will consider this case next.  Fig.~\ref{res2} shows the 
results for both cluster sizes with the same parameters as in 
Fig.~\ref{res1} but at half filling. Whereas the 1x1 cluster result 
displays the same features as in the doped case - enhanced density 
of states at the Fermi level accompanied by a parabolic minimum in 
the imaginary part of the self energy - the spectrum of the 2x2 
cluster calculation is completely different from the doped case: 
Instead of forming a quasi-particle resonance as in the DMFA, the
density of states develops a pseudogap at zero frequency and the
corresponding imaginary part of the self energy which is again 
almost isotropic displays a strongly enhanced scattering rate at the 
Fermi energy. This surprising and 
interesting behavior has two possible explanations. The first and
physically most appealing one is that short-ranged antiferromagnetic
fluctuations do indeed drive the system from a Fermi liquid into a 
non Fermi liquid at temperatures high compared to the N\'eel temperature. 
Note that the underlying mechanism is very similar to the interpretation 
of the pseudogaps observed in the high-$T_c$ compounds well above 
$T_c$. The second interpretation is that the nonlocal 
corrections yield a reduction in the critical value $U_c$ at which
the Mott-Hubbard metal-insulator transition occurs.

\begin{center}
\begin{figure}[h]
\epsfxsize=3.25in 
\epsffile{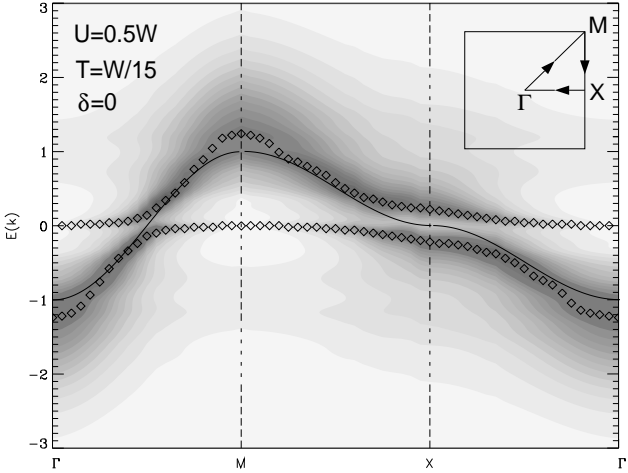}  

\vspace{0.5cm}
\caption{{Contour-plot of the spectral density for the 2x2 cluster 
calculation along the main symmetry directions as indicated in the 
inset for the same parameters as in Fig.~\ref{res2}. The dark color 
marks regions with high spectral density. The open symbols indicate 
the maxima of the spectral function. The solid line marks the dispersion 
for the noninteracting system.}}
\label{bs}
\end{figure}
\end{center}

A contour-plot of the spectral density $A({\bf k},\omega)$ obtained with
the bilinear interpolation scheme discussed earlier along the main symmetry
directions in the Brillouin zone is shown in Fig.~\ref{bs}. The dark shading 
marks regions with high spectral density. The open symbols in Fig.~\ref{bs} represent the positions of the most pronounced
local maxima of $A({\bf k},\omega)$ and can be viewed as effective band
structure of the 2D Hubbard model for the set of parameters under
consideration. Compared to the 
bandstructure of the noninteracting system (illustrated by the solid line) the 
interactions have various effects on the spectrum. The band of the 
noninteracting system splits into two separated bands above and below 
the Fermi-energy. Note that the spectral density corresponding to the 
nearly dispersionless features of the two bands is very low and comparatively broad in the regions 
without states of the noninteracting system. We again notice the opening of 
the pseudogap at the Fermi-energy which is most pronounced at the 
X-point $(\pi/2,\pi/2)$. But in addition to these effects we now can 
resolve additional incoherent background structures at the points 
$\Gamma$ at $E(k)\approx 1$ and M at $E(k)\approx -1$. These additional 
states are just shifted by the wave vector ${\bf Q}=(\pi,\pi)$ with 
respect to the main bandstructure.
\begin{center}
\begin{figure}[h]
\epsfxsize=3.25in 
\epsffile{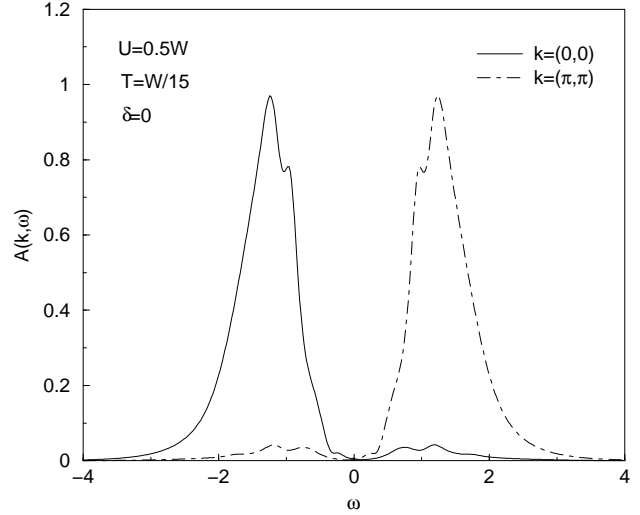}  

\vspace{0.5cm}
\caption{Spectral density at the points $\Gamma$ and M for the same 
parameters as in Fig.~\ref{res2} and \ref{bs}. Note the occurrence of 
shadow structures with small spectral weight in addition to the main 
structures.}
\label{bs2}
\end{figure}
\end{center}
In order to better resolve these structures we show the 
corresponding spectral density at the points $\Gamma$ and M in 
Fig.~\ref{bs2}. In addition to the main structure at $\omega<0$ for 
the point $\Gamma$ ($\omega>0$ for M) we notice satellites at 
$\omega>0$ for $\Gamma$ ($\omega<0$ for M) with small spectral weight. 
These new states are absent in the non-interacting system as well as
in the DMFA and result 
from the nonlocal antiferromagnetic correlations.  Even in the 
paramagnetic phase the short-ranged antiferromagnetic spin fluctuations 
are sufficient to produce this indication of the ordered phase. Such a
precursor effect of the antiferromagnetic long range order can for example be seen in the cuprates \cite{shadow} in Fermi-surface measurements. 
The observation of these spin-fluctuation-induced shadow states 
accompanied by an opening of a pseudogap strongly supports the first 
suggested scenario of the antiferromagnetic spin fluctuations driving 
the system to a non Fermi liquid.   
 
To gain more insight in the nature of the pseudogap and elucidate the 
physics of the observed non Fermi liquid behavior we added a next nearest
neighbor hopping $t^\prime$ to the hopping integrals $t_{ij}$ in the
Hamiltonian (\ref{HM}). The dispersion then becomes $\epsilon_{\bf
  k}=-2t(\cos k_x+\cos k_y)+4t^\prime\cos k_x\cos k_y$ and the Fermi
surface is no longer nested because $\epsilon_{{\bf k}+{\bf Q}}\neq
-\epsilon_{\bf k}$ for the nesting vector ${\bf Q}=(\pi,\pi)$. By
including $t^\prime$ we can thus frustrate the lattice and gradually
suppress antiferromagnetic spin fluctuations. On the other hand, since
we keep the non-interacting bandwidth $W$ and therefore the ratio
$U/W$ fixed when including the $t^\prime$, we do not expecte this
change to affect the Mott-Hubbard transition very
drastically. Therefore, if the pseudogap is a precursor of the
Mott-Hubbard transition we do not expect it to be influenced
dramatically  when we increase 
$t^\prime$. 
\begin{center}
\begin{figure}[h]
\epsfxsize=3.25in 
\epsffile{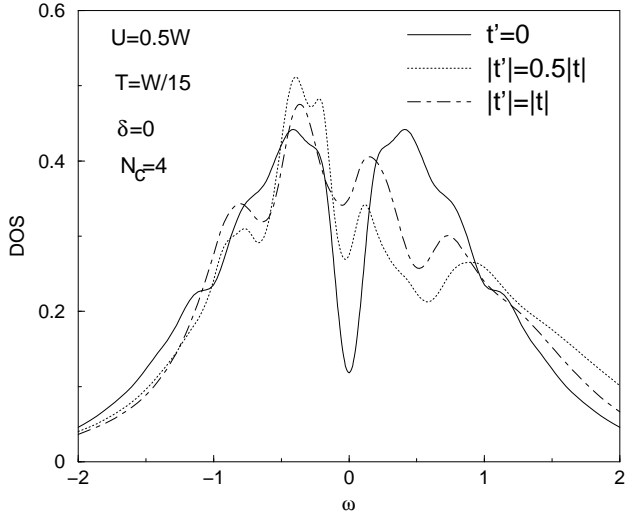}  
\caption{Local density of states for the 2x2 ($N_c=4$) cluster at fixed 
temperature $T=W/15$, interaction $U=0.5W$ and half filling for various 
values of the next nearest neighbor hopping integral $t^\prime$.}
\label{res3}
\end{figure}
\end{center}
Fig.~\ref{res3} shows the behavior of the local density of states as
the next nearest neighbor hopping $t^\prime$ and therefore the lattice
frustration increases. First note that we lose particle hole symmetry
for $t^\prime \neq 0$ because the bare density of states is no longer
particle hole symmetric. Obviously, as $t^\prime$ increases, the
pronounced pseudogap for $t^\prime=0$ gets gradually
reduced. However, even for the maximum value $|t^\prime|=|t|$ we still
see a small dip at zero frequency, which possibly can be attributed to 
the fact that even for a completely frustrated system extreme
short-ranged fluctuations will be present, which are however strongly
reduced in magnitude as compared to the nested situation. We thus
conclude that indeed the nonlocal, short-ranged
antiferromagnetic spin correlations are responsible for the
development of the pseudogap at the Fermi energy in the Hubbard model 
with $t^\prime=0$ at half filling, which in fact must be viewed as
precursor of an antiferromagnetically ordered state at much lower
temperatures. One highly interesting question to be addressed in our
future work will be of what precise nature this non Fermi liquid state
is and how it might be related to several phenomenological scenarios
proposed for the two-dimensional Hubbard model.

\begin{figure}[h]
\epsfxsize=3.6in 
\epsffile{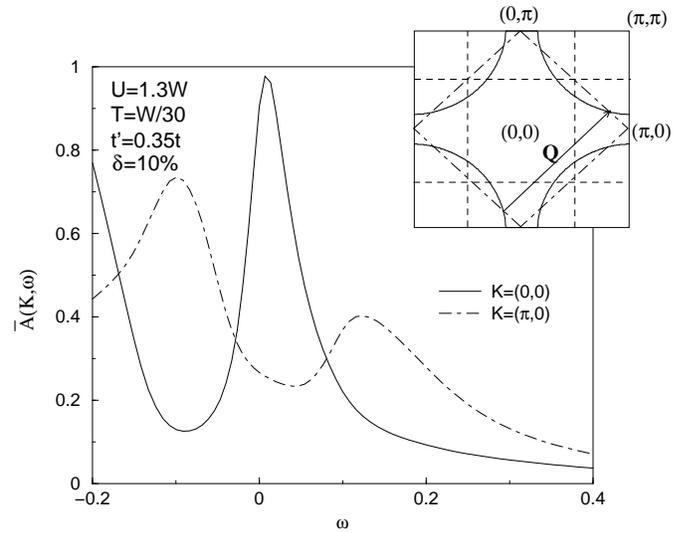}  
\caption{Coarse grained spectral function of the cells around ${\bf K}=(0,0)$ and $(\pi,0)$ for $N_c=4$ and a doping concentration 
$\delta=10\%$, temperature $T=W/30$ and hopping integrals $t=0.25$ and $t^\prime=-0.35t$. The inset shows the Fermisurface of the corresponding non-interacting system (solid line), the nested ${\bf k}$-points (dot-dashed line) and the boundaries of the coarse graining cells (dashed lines).}
\label{res4}
\end{figure}

As a further illustration of the ability of the DCA to include nonlocal 
correlations, we show results for a larger interaction strength $U=1.3W$ and lower temperature $T=W/30$. For the hopping integrals we chose $t=0.25$ and $t^\prime=-0.35t$ to immitate the measured Fermisurface of underdoped cuprates in the non-interacting system \cite{preformed0,schmalian}.  Since we expect the spectra to display sharp features at this low temperature, we refrain from performing a bilinear interpolation of the self-energy and show results for the coarse grained spectra only.  
Fig.~\ref{res4} shows the coarse grained spectral functions $\bar{A}({\bf K},\omega)=-1/\pi\Im m \bar{G}({\bf K},\omega)$ at ${\bf K}=(0,0)$ and ${\bf K}=(\pi,0)$ for a doping $\delta=10\%$ and temperature $T=W/30$ in a narrow region around the Fermienergy.         
First we notice a strong anisotropy in the coarse grained functions. For the states located in the cell around ${\bf K}=(0,0)$ the spectrum is peaked at the 
Fermi-energy and only slightly broadened due to the finite temperature 
as characteristic for a Fermi liquid. The situation 
is completely different in the cell around the point 
$(\pi,0)$. All the spectral weight is transfered to broad features at 
higher energies, which represents the incoherent part of the spectrum. 
Therefore a pseudogap opens at the Fermi-energy.

Our results are qualitatively in agreement with calculations \cite{schmalian} for the spin 
fluctuation model, in which the magnetic interaction between the 
quasiparticles is held responsible for the anomalous normal state properties 
of the cuprates. This method provides a direct explanation for the anisotropic 
behavior of the spectral density. Within this approach the electron-electron 
interaction is mediated by an imperially determined dynamical spin 
susceptibility. Since this susceptibility is strongly peaked at the 
antiferromagnetic wave vector ${\bf Q}=(\pi,\pi)$, one has to distinguish 
two different regions of the Fermi surface: Quasiparticles in regions of the Fermisurface which can be connected by the wave vector $\bf Q$ are 
called {\em hot quasiparticles} because they feel the full effects of the 
spin fluctuation induced interaction because of nesting. This is illustrated in the inset of Fig.~\ref{res4}, which displays the corresponding non-interacting Fermisurface for the chosen parameters. One can notice that the hot quasiparticles are located in the cell around $(\pi,0)$ and therefore represented by the coarse grained spectral function at ${\bf K}=(\pi,0)$. On the other hand the {\em cold 
quasiparticles} located along the diagonal couple only weakly to 
the spin excitations, since the Fermisurface in this region is not nested. This part of the Femisurface falls in the cell around ${\bf K}=(0,0)$ and therefore the spectrum in this cell displays Fermi-liquid like behavior. The spectrum around $(\pi,0)$ on the other hand gets strongly 
renormalized due to the strong coupling to the spin excitations. This 
phenomenological picture provides a direct explanation of our results 
qualitatively consistent with the calculations for the spin fluctuation 
model.

Calculations for larger doping ($\delta=20\%$) (not shown here) show that 
this effect of the anisotropic behavior of the spectrum and the opening of 
the pseudogap in the hot regions disappears. This observation can also be 
understood within the picture of the spin fluctuation induced correlations, 
since the antiferromagnetic spin fluctuations become strongly suppressed upon 
doping.                

\section{Conclusion}

We motivated the recently introduced Dynamical cluster approximation (DCA) 
by its microscopic definition based on a choice for the Laue function.
It partially restores the momentum conservation at the internal 
vertices which was relinquished in the Dynamical Mean Field 
Approximation (DMFA).  The resulting theory maps the lattice problem onto a 
self-consistently embedded periodic cluster of size $N_c$.  The DCA is a 
fully causal and systematic approximation to the full lattice problem
with corrections ${\cal{O}}(1/N_c)$ in two dimensions. We develop a Non 
Crossing Approximation (NCA) to solve the effective cluster problem which 
is a systematic $\Phi$-derivable approximation to the cluster problem 
with corrections ${\cal{O}}(1/N_c^3)$. 

        We applied our DCA-NCA formalism to the Hubbard model on a square 
lattice and calculated the single-particle properties when $N_c=1$ and $N_c=4$. 
For a highly doped system, with $\delta=15\%$ and a Hubbard $U$ of half the 
bare bandwidth, nonlocal correlations present when $N_c=4$ turned out to be 
unimportant and both cluster sizes yield qualitatively similar results with 
an enhanced density of states and a minimum in the scattering rate at the 
Fermi energy. Thus, independent of the cluster size, the highly doped system 
showed Fermi liquid character. However, as the doping decreases, the 
non-local correlations become far more important, and the half-filled 
system displays strikingly different results for the two cluster sizes. 
Whereas the $N_c=1$ cluster result still displays Fermi liquid like behavior, 
the $N_c=4$ cluster calculation shows the opening of a pseudogap in the 
density of states and therefore non Fermi liquid character. Calculations with 
a next nearest neighbor hopping $t^\prime$ show evidence that this pseudogap 
is due to antiferromagnetic spin correlations and therefore a single particle 
precursor of the antiferromagnetic phase transition. The pseudogap persists upon weak doping in qualitative agreement with the spin fluctuation model for the cuprates.

Hence our calculations have shown that for the weakly doped system, nonlocal 
correlations play an important role on the single particle properties and 
change the character of the system from a Fermi liquid to a non Fermi liquid. 
These non-local features are missing in the DMFA spectra, but appear in
the DCA spectra as soon as the cluster size exceeds one.

Acknowledgements: It is a pleasure to acknowledge discussions with 
P.G.J.\ van Dongen, 
D.\ Hess, 
M.\ Hettler,
H.R.\ Krishnamurthy,
E.\ M\"uller-Hartmann, 
and F.-C. Zhang.
This work was supported by NSF grants DMR-9704021, DMR-9357199, 
the Graduiertenkolleg ``Kom\-plexi\-t\"at in Festk\"orpern'' and the NATO
Collaborative Research Grant CRG970311.
Computer support was provided by the Ohio Supercomputer Center and the 
Leibnitz-Rech\-en\-zen\-trum, Munich.

\appendix

\section{On the analyticity of $\Gamma({\bf K},z)$\label{app}}
\label{app:a}

The following proof of the analyticity of $\Gamma({\bf K},z)$ is based on
the derivation of an analytic
expression for the cluster Green's function
$$
\bar{G}({\bf K},z)=\frac{N_c}{N}\sum\limits_{{\bf
k}'}\frac{1}{z+\mu-\epsilon_{{\bf K}+{\bf k}'}-\Sigma_c({\bf K},z)}\;\;.
$$
The self-energy function $\Sigma_c({\bf K},z)$ is assumed to be analytic 
in the upper and lower half of the complex plane with 
${\rm sign}(\Im m\Sigma_c({\bf K},z))=-{\rm sign}\Im m(z)$.
Our proof employs standard methods of projection technique \cite{grosso,magnus}.
To this end we abbreviate $\xiK=z+\mu-\Sigma_c({\bf K},z)$, $\Im m(\xiK)>0$
and introduce
a $N/N_c$ dimensional set of linearly independent vectors $\ket{{\bf k}'}$ and
a linear hermitian operator $\HK$ satisfying
$\HK\ket{{\bf k}'}=\epsilon_{{\bf K}+{\bf k}'}\ket{{\bf k}'}$ and
$\sum\limits_{{\bf k}'}\ket{{\bf k}'}\bra{{\bf k}'}=1$. There obviously exists
another vector
$\ket{f_0}$ with $\bra{{\bf k}'}f_0\rangle=\sqrt{N_c/N}$ for all ${\bf k}'$.
With these conventions we may
write
$$
\bar{G}({\bf K},z)=\bra{f_0}\frac{1}{\xiK-\HK}\ket{f_0}\;\;.
$$
The resolvent operator may be trivially rewritten as
$$
\frac{1}{\xiK-\HK}=\frac{1}{\xiK}+\frac{1}{\xiK}\HK
\frac{1}{\xiK-\HK}
$$
and thus
$$
\bra{f_0}\frac{1}{\xiK-\HK}\ket{f_0} = \frac{1}{\xiK} + \frac{1}{\xiK}
\bra{f_0}\HK\frac{1}{\xiK-\HK}\ket{f_0}\;\;,
$$
where we made use of $\bracket{f_0}{f_0}=1$. We now define two projection
operators
$P=\ket{f_0}\bra{f_0}$ and $Q=1-P$ and insert $1=P+Q$ after ${\cal H}_{\bf K}$
in the second term,
leading to
$$
\bra{f_0}\frac{1}{\xiK-\HK}\ket{f_0} = \begin{array}[t]{l}
\displaystyle\frac{1}{\xiK} +
\frac{1}{\xiK}\bra{f_0}\HK\ket{f_0}\bra{f_0}\frac{1}{\xiK-\HK}\ket{f_0}\\[5mm]
\displaystyle+ \frac{1}{\xiK}\bra{f_0}\HK Q\frac{1}{\xiK-\HK}\ket{f_0}\;\;.
\end{array}
$$
Since furthermore
$$
\frac{1}{\xiK-\HK}=\begin{array}[t]{l}
\displaystyle\frac{1}{\xiK-\HK Q-\HK P}=
\displaystyle\frac{1}{\xiK-\HK Q}+\\[5mm]
\displaystyle\frac{1}{\xiK-\HK Q}\HK P\frac{1}{\xiK-\HK}
\end{array}
$$
and due to $Q\ket{f_0}=0$
$$
Q\frac{1}{\xiK-\HK Q}\ket{f_0}=0
$$
we finally obtain
$$
\begin{array}[t]{l}
\displaystyle\bra{f_0}\frac{1}{\xiK-\HK}\ket{f_0} =
\displaystyle \frac{1}{\xiK}+
\frac{1}{\xiK}\bra{f_0}\HK\ket{f_0}\bra{f_0}\frac{1}{\xiK-\HK}\ket{f_0}\\[5mm]
\displaystyle +\frac{1}{\xiK}\bra{f_0}\HK Q\frac{1}{\xiK-\HK Q}\HK\ket{f_0}
\bra{f_0}\frac{1}{\xiK-\HK}\ket{f_0}\;\;.
\end{array}
$$
With $Q^2=Q$ we may rewrite
$$
\HK Q\frac{1}{\xiK-\HK Q}\HK = \HK Q\frac{1}{\xiK-Q\HK Q}Q\HK\;\;.
$$
With the abbreviations 
$$
\Gamma({\bf K},z) = \bra{f_0}\HK Q\frac{1}{\xiK-Q\HK Q}Q\HK\ket{f_0}
$$
and
$$
\bar{\epsilon}_{\bf K}=\bra{f_0}\HK\ket{f_0}=\frac{N_c}{N}\sum\limits_{{\bf
k}'}\epsilon_{{\bf K}+{\bf k}'}
$$
the final result is
\begin{equation}\label{app_Gbar}
\bar{G}({\bf K},z) = \frac{1}{\xiK-\bar{\epsilon}_{\bf K}-\Gamma({\bf
K},z)}\;\;.
\end{equation}
Note however that the
averaging procedure replaces the kinetic energy of the lattice $\epsilon_{\bf
k}$ by a quantity coarse grained
onto the cluster.

We are now left with the proof that ${\rm sign}\Im m(\Gamma({\bf K},z))=-{\rm
sign}\Im m(z)$ is fulfilled.
This however can easily be done by repeating the above step for the new vector
$\ket{f_1}=Q\HK\ket{f_0}=\HK\ket{f_0}-\bar{\epsilon}_{\bf K}\ket{f_0}$,
$\bracket{f_0}{f_1}=0$, appearing in the definition of $\Gamma({\bf K},z)$.
Note that this is simply
the first step in a Schmitt-Graham procedure to generate an orthogonal set of
vectors. With $\tilde{\cal H}_{\bf K}
=Q\HK Q$ it follows
\begin{equation}\label{Gamma}
\Gamma({\bf K},z)=\frac{b_0^2}{\xiK-a_1-\bar{\Gamma}({\bf K},z)}\;\;,
\end{equation}
where
$$
\begin{array}{l}
\displaystyle b_0^2=\frac{\bracket{f_1}{f_1}}{\bracket{f_0}{f_0}}\ge0\\[5mm]
\displaystyle a_1 = \frac{\bra{f_1}\HK\ket{f_1}}{\bracket{f_1}{f_1}}\\[5mm]
\displaystyle \bar{\Gamma}({\bf K},z)=\frac{1}{\bracket{f_1}{f_1}}
\bra{f_1}\tilde{\cal H}_{\bf K}\tilde{Q}\frac{1}{\xiK-\tilde{Q}\tilde{\cal
H}_{\bf K}\tilde{Q}}
\tilde{Q}\tilde{\cal H}_{\bf K}\ket{f_1}\;\;,
\end{array}
$$
where $\tilde{Q}$ now projects onto the subspace orthogonal to $\ket{f_0}$ and
$\ket{f_1}$.
It is clear from the above result that this procedure can be repeated, leading
to a sequence of mutually orthogonal
vectors
$\ket{f_n}=\HK\ket{f_{n-1}}-a_{n-1}\ket{f_{n-1}}-b_{n-2}^2\ket{f_{n-2}}$ and a
continued fraction representation of $\bar{G}({\bf K},z)$ with coefficients
$$
\begin{array}{l}
\displaystyle
b_{n-1}^2=\frac{\bracket{f_n}{f_n}}{\bracket{f_{n-1}}{f_{n-1}}}\ge0\\[5mm]
\displaystyle a_n=\frac{\bra{f_n}\HK\ket{f_n}}{\bracket{f_n}{f_n}}
\end{array}
$$
for $n\ge1$.
It is important to emphasize that the resulting coefficients $b_{n-1}^2$ are
non-negative by construction.
This however immediately leads to the desired relation ${\rm sign}\Im
m(\Gamma({\bf K},z))=-{\rm sign}\Im m(z)$ and
hence to the causality of $\Gamma({\bf K},z)$.

\section{Proof of $\Gamma({\bf K})\sim{\cal O}(1/N_c)$}
\label{app:b}
The following proof is based on the definitions (\ref{G_bar_sum},\ref{G_bar_loc}) of the coarse grained Green function $\bar{G}$. With $g^{-1}({\bf K})=\omega-\bar{\epsilon}_{\bf K}-\bar{\Sigma}({\bf K})$ we can rewrite $\bar{G}$ in the form 
\begin{eqnarray}
\bar{G}({\bf K})&=&\frac{N_c}{N}\sum_{\tilde{\bf k}}G({\bf K}+\tilde{\bf k})\nonumber\\
&=&\frac{1}{g^{-1}({\bf K})-\Gamma({\bf K})}\quad\mbox{,}
\label{Gg}
\end{eqnarray} 
where we dropped the frequency argument for convenience. By defining
\begin{equation}
t_{{\bf K}+\tilde{\bf k}}=\epsilon_{{\bf K}+\tilde{\bf k}}-\bar{\epsilon}_{\bf K}\quad\mbox{,}
\end{equation} 
such that $N_c/N \sum_{\tilde{\bf k}}t_{{\bf K}+\tilde{\bf k}}=0$ we can make use of the exact relation
\begin{equation}
G({\bf K}+\tilde{\bf k})=g({\bf K})+g({\bf K})t_{{\bf K}+\tilde{\bf k}}G({\bf K}+\tilde{\bf k})\quad\mbox{.}
\label{Dyson}
\end{equation}
By inserting (\ref{Dyson}) in (\ref{Gg}) it is straightforward to show that $\Gamma({\bf K})$ is given by
\begin{equation}
\Gamma({\bf K})=\frac{\displaystyle \frac{N_c}{N}\sum_{\tilde{\bf k}}t^2_{{\bf K}+\tilde{\bf k}}G({\bf K}+\tilde{\bf k})}{\displaystyle 1+\frac{N_c}{N}\sum_{\tilde{\bf k}}t_{{\bf K}+\tilde{\bf k}}G({\bf K}+\tilde{\bf k})}\quad\mbox{.}
\label{GN}
\end{equation}  
By performing a Taylor series expansion of $t_{{\bf K}+\tilde{\bf k}}$ around the cluster points $\bf K$ it can be seen that $t_{{\bf K}+\tilde{\bf k}}$ is of order ${\cal O}(\Delta k)$, where $\Delta k=2\pi/N_c^{1/D}$. Therefore we see from Eq.~\ref{GN} that $\Gamma({\bf K})$ is of order ${\cal O}((\Delta k)^2)$ and for two dimensions we finally get the result
\begin{equation}  
\Gamma({\bf K})\sim{\cal O}(1/N_c)\quad\mbox{.}
\end{equation}


\begin{thebibliography}{}
\bibitem{ED} It is possible that an exact-diagonalization based generalization 
of the approach presented here will make larger clusters more tractable.  The 
first step in our NCA approach involves diagonalizing the molecular 
cluster. If this was the limiting step in our technique, we should be
able to treat clusters as large as those presently studied by exact 
diagonalization.  However, the remaining numerical task of calculating the 
coupled equations (\ref{P_m}) and (\ref{self_m}) self-consistently is 
necessary for the NCA to be $\Phi$ derivable, but it becomes formidable 
as the cluster size increases.  The $\Phi$-derivability is important since 
it guarantees that our results will be consistent at the one and two-particle 
level.  If we were to relax this condition, and neglect the higher-order,
${\cal{O}}(1/N_c^3)$ and higher, non-crossing diagrams by using bare 
resolvents in (\ref{self_m}), then we could treat significantly larger 
clusters and the approximation would still be correct up to ${\cal O}(1/N_c)$.
\end{thebibliography}
\end{document}